\documentclass[printer]{aa}  
\usepackage{graphicx}
\usepackage{txfonts}
\begin{document}
   \title{Kinematic segregation of nearby  disk stars from the Hipparcos database }

   \author{R. E. de Souza
          \inst{1}
          \and
          R. Teixeira\inst{1}
          }
   \titlerunning{Kinematics of Disk Stars}

   \offprints{R. E. de Souza}

   \institute{Dept. Astronomia, IAG/USP, University of S\~ao Paulo,
   Rua do Mat\~ao 1226, 05508-900   S\~ao Paulo -SP, Brazil}

   \date{Received August 16, 2006; accepted March 20, 2007}

% \abstract{}{}{}{}{} 
% 5 {} token are mandatory
 
  \abstract
     {}
   {To better understand our Galaxy, we investigate the pertinency of describing the system of nearby disk stars in terms of a two-components Schwarzschild velocity distribution.Using the proper motion and parallax information of Hipparcos database, we determine the parameters characterizing the local stellar velocity field of a sample of 22000 disk stars. The sample we use is essentially the same as the one described by the criteria adopted to study the LSR and the stream motion of the nearby stellar population.}
   { The selected data is modeled with a two component Schwarzschild velocity distribution whose parameters were determined by a least-square regression. The celestial sphere was divided into 72 equal area regions used to determine the parameters minimizing the final velocity distribution function.}
   {We verify that the results are not significantly different for the early type stars from the classical treatment using a single Gaussian population distribution. For late type stars in the subgiant branch, in contrast, we verify that the two-component model gives a much more satisfactory representation. Our results indicate for each spectral class the presence of these late type stars  of a low velocity dispersion population, with $\sigma_u \simeq 20 {\rm km.s}^{-1}$  coexisting with a high velocity dispersion  population having  $\sigma_u \simeq 40 {\rm km.s}^{-1}$. Both populations belong to the disk with scale heights of $120 {\rm pc}$ and $220 {\rm pc}$, respectively, relative to the Galactic disk.}
   {}


   \maketitle
%
%________________________________________________________________

\section{Introduction}

The kinematics of nearby disk stars is a longstanding issue in the literature has led to some major discoveries in  the Galactic structure (see North \cite{North}  for a complete review in this subject). It is clearly important to properly characterize the motion of disk stars in the solar neighborhood. Most of our present knowledge of the dynamical properties, evolution, and origin of the Galactic disk depends on analyzing of the local stars. Hopefully this knowledge will also improve our understanding of the nearby galaxies where a similar star population is far more difficult to observe directly. A large body of information is already at our disposal to support our current knowledge of the disk (see for instance Binney \& Merrifield \cite{BinMerr} and Binney \& Tremaine \cite{BinTrem} for a fully review of this subject). 

Normally the velocity field in the solar neighborhood is described by a single Schwarzschild (\cite{Schwarzs}) velocity distribution, which is known to obey the collisionless Boltzmann equation. Due to the non-collisional nature of the star fluid, the velocity distribution can sustain different dispersions along the three Galactic coordinate axes. But this distribution does not remain strictly constant since the individual velocities of stars gradually change on the Hubble time scale as a response to the different evolutionary processes involved. Therefore by studying the statistics of the presently observed velocities in the solar region, it is possible to infer the dispersions of the velocity distribution and hopefully to see how they have changed with age, metallicity, and spectral class of the stellar population as proposed by many authors (Str\"omgren \cite{Stromgren87}; Soubiran and Girard \cite{Soubiran05} and references therein).   

A large body of data selected by age, metallicity, and spectral type attest to the validity of this bold description. One interesting issue that remains is how well we can separate the different kinematical populations of disk stars. It is a settled matter that disk stars of a given spectroscopic family share the same gross kinematical properties. Young early type stars  are more concentrated towards the Galactic plane having a lower velocity dispersion in the range 10-20 ${\rm km.s}^{-1}$. In contrast late type stars, dominated by an older population, have a higher velocity dispersion (30-40 ${\rm km.s}^{-1}$) and are more scattered around the Galactic plane (see Fig. \ref{DensZType}). 

The same basic picture emerges when we sample nearby stars selected by their ages or metallicity. The young metal-rich population tends to have the lower velocity dispersion typical of the thin disk, while in the old metal-poor stars its value is definitely higher. These two large groups of stars also share the same velocity anisotropies, as measured by the ratio of the velocity dispersions along the radial direction ($\sigma_U$) and in the transversal direction ( both in the Galactic rotation direction, $\sigma_V$ , and perpendicular to the plane of the disk, $\sigma_W$ ). 

A question that is also important in this context is whether an age spread should remain  inside a given stellar population. Therefore we should expect that the young and the old stars of a given stellar type group should contribute differently to the overall dispersion velocities. Clearly for early type stars, the age spread is not expected to be large since these are short-lived objects, so we should also expect their kinematical properties to be more homogenous. But for late type stars we should be able to observe a much larger age spread, and therefore their global kinematics should be much more complex and sensitive to the age composition of the population. We should expect for these objects a population of younger newly-born late type stars, therefore sharing the same kinematical properties as thin disk young stars, superposed on an old population of late type stars that have already suffered the cumulative effect of living inside a disk with gravitational potential irregularities. An alternative to this long-term, small gradual changing scenario was proposed by Carney et al. (\cite{Carney89}) based on possible detection of a discrete population of stars having [m/H] $\simeq -0.5$ and a velocity dispersion of $40 {\rm km.s}^{-1}$ perpendicular to the disk. The existence of such a discrete population argues in favor of a relic population resulting from an early merger event occurring shortly after the formation of the disk. 

An earlier attempt to detect the effects of such a superposition of two different kinematical populations has already been done by Oort (\cite{Oort32}), who used a relatively small sample of objects to conclude that this might be the case for late K-M dwarfs. According to that author, the kinematics of the two populations have $\sigma_w= 10 {\rm km.s}^{-1}$ and $\sigma_w=25 {\rm km.s}^{-1}$. In this paper we focus our attention on detecting a second kinematical population using the proper-motion and parallax data of a sample of stars in the Hipparcos catalogue (ESA 1997). That is the best database available to date for that kind of investigation because it provides positions, parallaxes, and annual proper motions for about 120 000 stars with an astrometric precision of $1 mas$, allowing a discussion of this subject to a level of accuracy never attained before. Although the available radial velocity data  is not complete, we can still use the homogeneous set of proper-motion and parallax observations to deduce the Galactic latitude and longitude components of the velocity field and constrain the kinematical description with that information. In other words, even if we cannot determine the individual components (U, V, W) for each star, we can develop a formalism to statistically determine the velocity dispersions $\sigma_U$, $\sigma_V$, and $\sigma_W$ for the whole population using this large sample of objects.

Among many others authors, Mignard (2000) has used this same approach to study the local kinematics. He constructed the Schwarchild probability distribution based on a selected set of approximately 22 000 Hipparcos stars by deducing the parameters that characterize their local kinematic properties divided by spectroscopic groups. In the present work we use the same sample as Mignard(2000) to discuss another possibility for representing the probability distribution of velocities. Instead of the single Gaussian  normally used to represent the Schwarzchild distribution, we used a two-Gaussian expression representing a two-fluid population. We used statistical tests to verify which of these descriptions gives a better representation of the real data. We present in  Sect. 2 a discussion of the selection criteria to define a homogeneous sample of the nearby stellar population using the Hipparcos database. In Sect. 3 we present the basic description adopted to the local velocity field showing how the parameters of this distribution changes with the Galactic coordinate orientation. In Sect. 4 we discuss the kinematical parameters obtained in this work. Section 5 contains our main conclusions and prospects for future work.  

%__________________________________________________________________

\section{Material}

In Mignard (2000) we can find a very judicious selection of Hipparcos stars suitable for studying and discussing the local kinematics. He based his selection on five criteria that assert a sample of Hipparcos stars well-distributed in the celestial sphere and in spectral classes in a way that avoids any statistical biases (see Table \ref{tab0} and Fig. \ref{MapStar}). 

\begin{table}[ht]
  \caption{Distribution of selected stars by spectral type interval estimated from the color index.}\label{tab0}
\centerline{
  \begin{tabular}{ccc}
     \hline
    B-V & spec. type & N$_{star}$ \\
      \hline
      \hline
 0.00-0.15 & A0-A5 & 4202 \\
 0.15-0.30 & A5-F0 & 3185 \\
 0.30-0.45 & F0-F5 & 2837 \\
 0.85-1.15 & K0-K5 & 6533 \\
 1.15-1.40 & K5-M0 & 3350 \\
 1.40-1.60 & M0-M5 & 2285 \\
     \hline
     \hline
  \end{tabular}
  }
\end{table}

\begin{figure*}
 \centering
 \includegraphics[width=\textwidth]{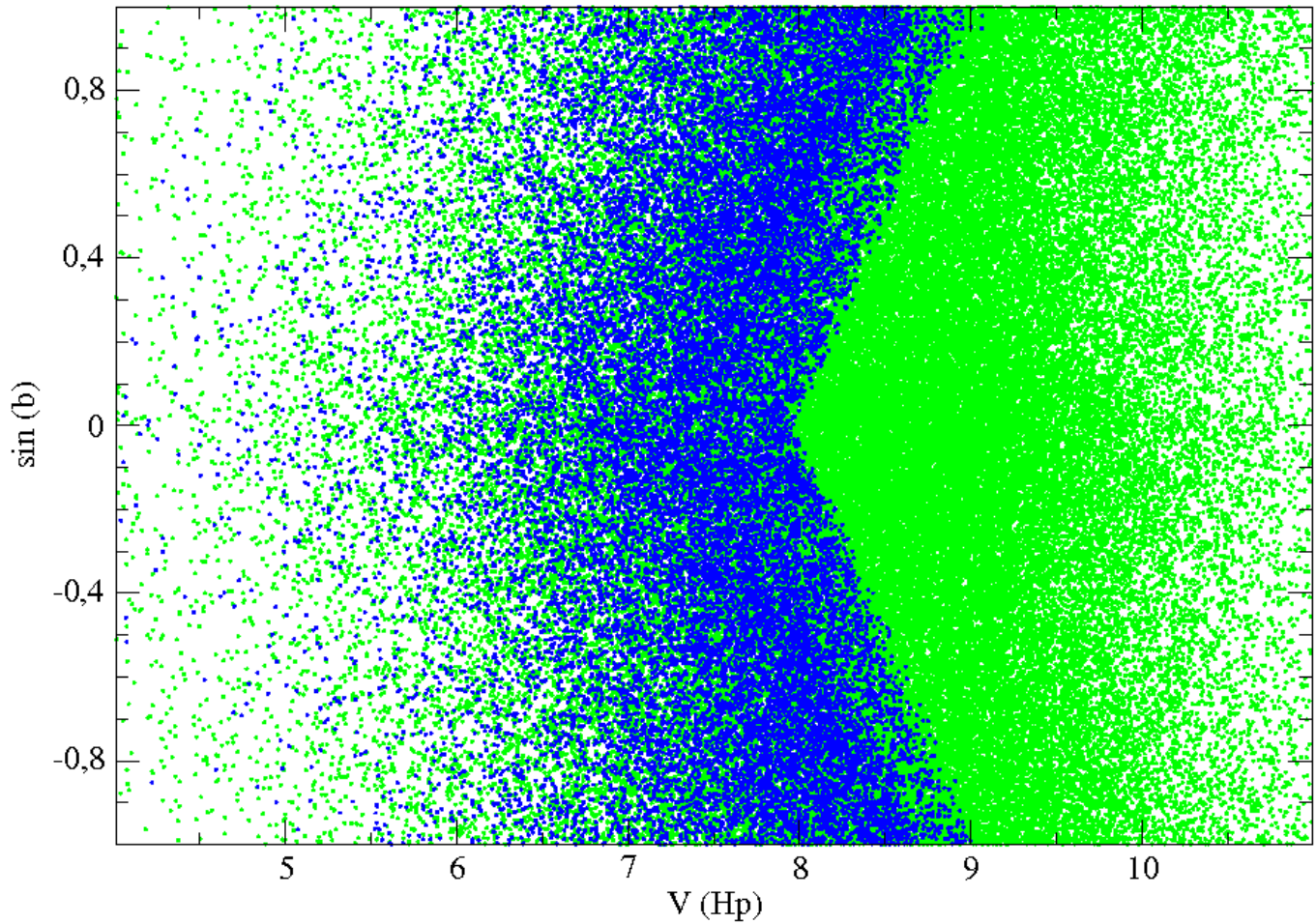}
 \caption{Galactic latitude distribution of selected objects (blue points) and the  rejected Hipparcos stars (green points). We can see  the sharp contour corresponding to the 70\% completeness limit in relation to the Tycho catalogue.}\label{LatMag}
\end{figure*}

The Mignard criteria, which we also adopted for selecting our sample from the Hipparcos catalogue, are summarized as: a) only single stars, due to their better astrometric mean precision and better definition of their spectral type; b) stars lying between 0.1 to 2.0 Kpc. The closer objects were eliminated due to their strong influence on the  solution of the solar motion. In this selected range of distance, some of the individual parallaxes might be not very precise, but we notice that even in this case their statistical meaning remains reliable for this study; c) through a detailed study of the completeness of Hipparcos stars, based on the more dense Tycho star catalogue (ESA 1997), a completeness limit larger than $70\%$ was established by a criteria using Galactic latitude, magnitude, and color; d) the stars were grouped in gross spectral types based on their colors; and e) stars with velocities higher than  $60-90 {\rm km.s}^{-1}$, depending on the spectral class, were excluded from the statistical analysis. We retain the first four criteria in our sample, but for the last one adopt a less restrictive criterion discarding only those stars with peculiar velocity higher than $100 {\rm.s}^{-1}$. This would correspond approximately to a three sigma level cut using the higher limit expectation for the disk velocity dispersion found in the literature ($\simeq 30 {\rm km.s}^{-1}$), so we feel confident that the high-velocity wing was faithfully sampled in our study.

\begin{figure*}
 \centering
 \includegraphics[width=\textwidth]{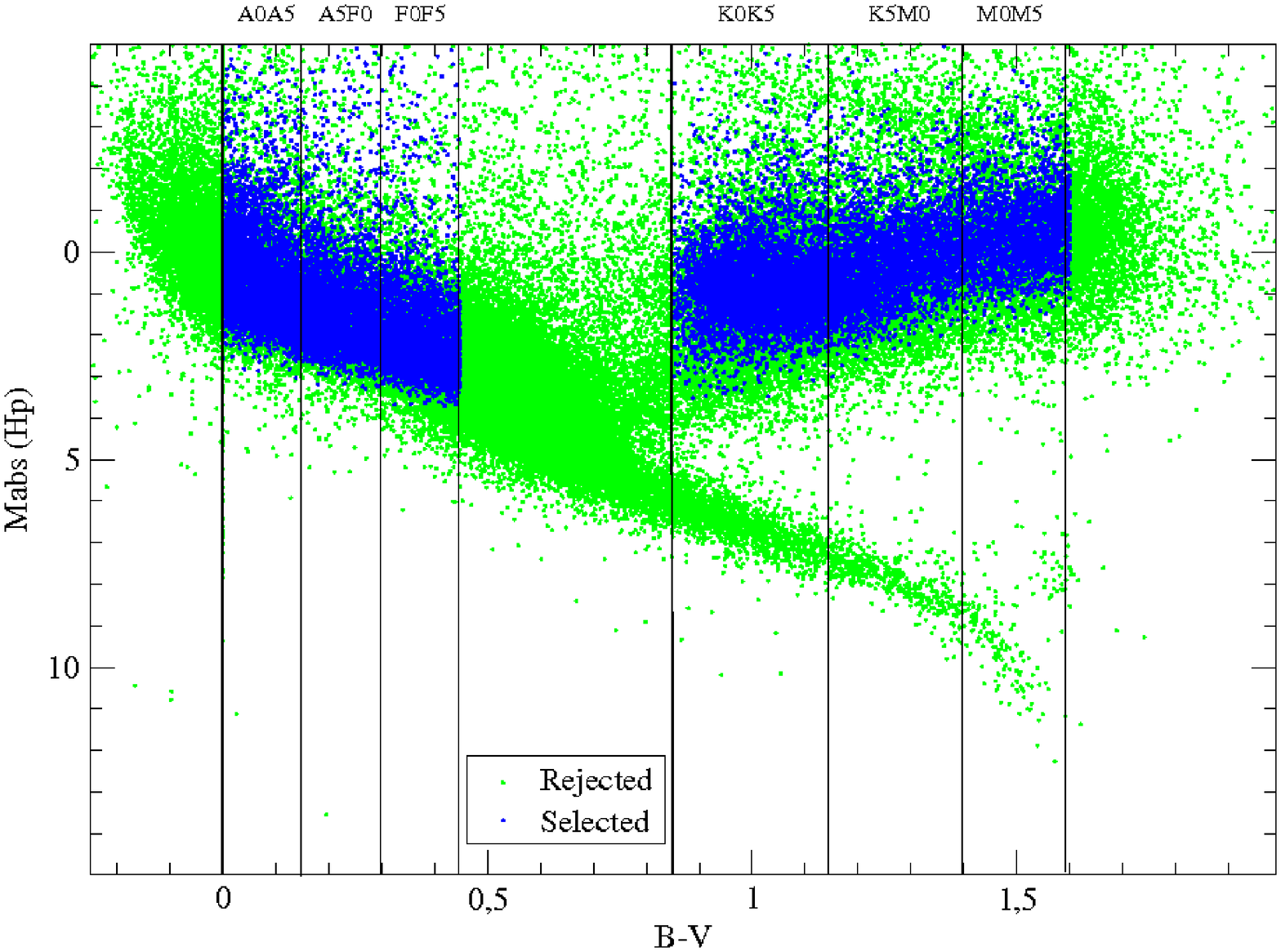}
 \caption{HR diagram of the selected (blue points) and all Hipparcos stars (green points).}\label{HRDiag}
\end{figure*}

In Fig. \ref{LatMag} we present the latitude distribution of the sample selected by our criteria. The sharp contour of the selected objects corresponds to the 70\% completeness limit derived from a comparison with the Tycho catalog. In Table \ref{tab0} we present the distribution of the selected stars as a function of their spectral group. The sample is essentially identical to the one used by Mignard except for minor differences due to the fact that we adopted an analytical criterion describing the limiting magnitude contour ( $V_{lim}=7.68+1.15| (\sin(b)|-0.43[(B-V)-0.72]$ ) to mimic the selection criteria based on the completeness. In particular, the color term was introduced to reproduce the variation in the completeness  limit with spectral type. The effect of the selection criteria on the HR diagram of our sample can be examined in Fig. \ref{HRDiag}.  We can see that the early type objects in our sample are basically populated by young-disk main sequence objects, and we exclude the very young ones that have peculiar kinematics associated to the region of their birth. In the case of late type objects the adopted criteria basically select  subgiants and giants. All the lower main sequence objects were excluded by adopting the limiting distance and the magnitude detection limit of the satellite. The intermediate color objects ($0.45<B-V<0.85$) were also excluded due to the difficulty of assessing an spectroscopic type based only on the color information. In this region, corresponding to the F5-K0 interval, the population is mixed by the presence of both dwarfs and giant stars. 

\begin{figure*}
 \centering
 \includegraphics[width=\textwidth]{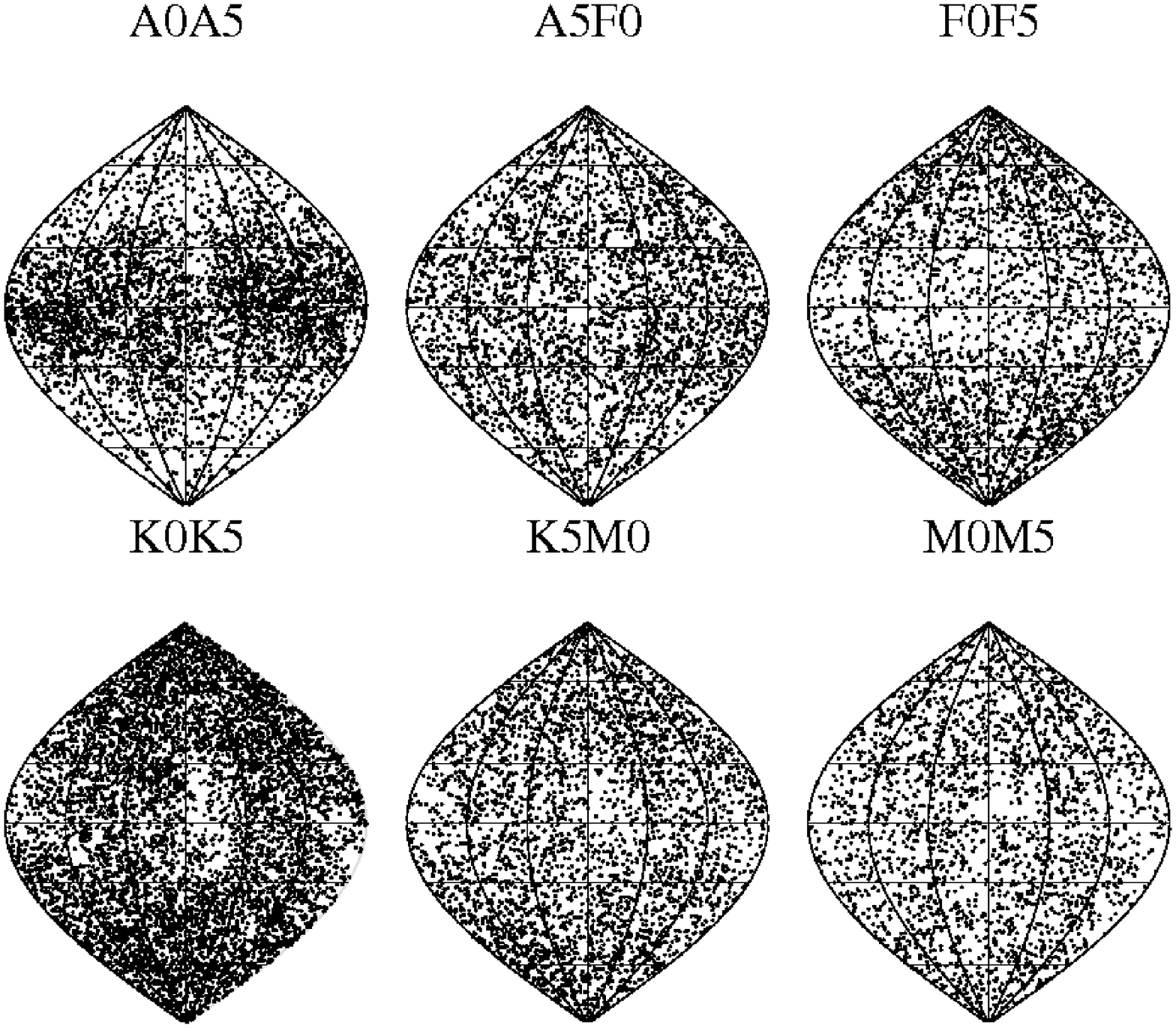}
 \caption{Distribution of each spectral type in the celestial sphere. Early type stars are clearly more concentrated towards the Galactic plane while the late types are more evenly distributed.}\label{MapStar}
\end{figure*}

\begin{figure*}
 \centering
 \includegraphics[width=\textwidth]{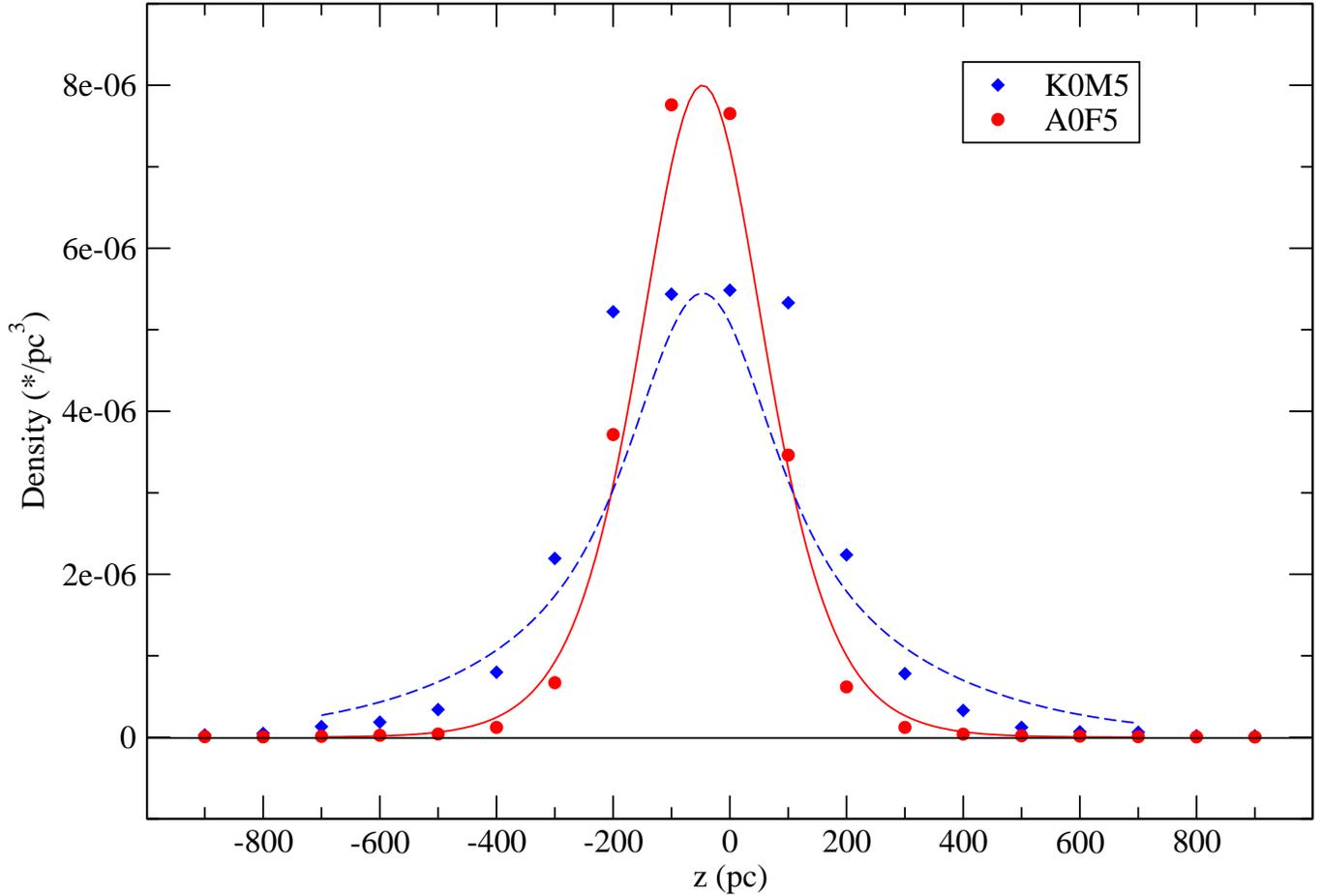}
 \caption{Density of early and late type stars as a function of the distance to the galactic plane. The two curves represent the expected one component isothermal disk density distribution ($\rho=\rho_{d0}{\rm sech^2}(z/z_0)$) with $z_0=145.3$ pc for early type stars (red continous line). For the late type star, the blue dashed line represents the expected composition of the two population having $z_0=145.3$pc and $z_0=396.3 $ as explained in Sect. 4.}\label{DensZType}
\end{figure*}

The distribution of the selected objects in the celestial sphere is presented in Fig. \ref{MapStar}, as expected, showing that early type stars are clearly more concentrated toward the plane of the Galactic disk. In contrast, late type giant stars are more uniformly distributed due to their larger scale height. This effect is more clearly seen in Fig. \ref{DensZType} where we show the density of the distribution as a function of the vertical distance  to the galactic plane. To obtain this density profile, we cut the selected sample into intervals of 100 pc of height above the Galactic plane counting all objects inside each of those elementary cylinders. Early type stars have been combined into a single class A0-F5 and are tightly distributed around the plane of the disk. Late type objects have been combined into a single K0-M5 class and are clearly distributed over larger distances. For a single isothermal population of an autogravitating disk, we should expect that this density profile follow the relation $\rho_d(z)=\rho_{d0} {\rm sech}^2 (z/z_0)\simeq \rho_{d0}\exp(-z^2/2z_0^2)$, where $z_0$ is the scale length and $\rho_{d0}$ represents the mass density in plane of the disk (Spitzer \cite{Spitzer42}). At moderate distances from the plane, this profile is approximately described by a Gaussian distribution. The two continuous lines presented in Fig. \ref{DensZType} follow this density law and do not intend to really be a fit to the observed density distribution, but simply to reflect the effect of the kinematic properties determined in this work (see more details in Sects. 3 and 4) and serve to illustrate the scale length of these two populations. 

We can see that early type objects are relatively close to the thin disk, and their dispersion around the Galactic disk determined directly from the density profile is $\sigma_z=124 {\rm pc}$.  The density profile of late type objects, on the contrary, is clearly spread out over larger distances as indicated by the height dispersion of $\sigma_z=210 {\rm pc}$. Nevertheless, both populations are clearly concentrated inside the so-called Galactic thin disk as estimated by other authors ($|z|< 350 {\rm pc}$) (see for example Soubiran et al. \cite{Soubiran03}). We note that the thick disk component would be visible only at larger distances from the Galactic plane ($|z| \simeq 1 {\rm kpc}$) and is not sampled in our data, although there might be some degree of contamination. The point of maximum density on both curves presented in Fig. \ref{DensZType} does not correspond exactly to $z=0$, meaning that the Sun is slightly displaced relative to the Galactic plane.

\section{Formalism}

The velocity vector of any star referred to the galactic center can be described as a composition of the local galactic rotation ($\vec{\rm V(R)}$) and a peculiar velocity ($\vec{\rm v}_*$) related to the individual orbit of that star in the galactic gravitational potential. The particular value of these two vector fields depends on the stellar population we are considering. Since these two components also affect the solar motion definition, it follows that the observed motion of any star relative to the Sun (${\vec{\rm\tilde v}_*}$) is also dependent on the stellar group considered

\begin{equation}
{\vec{\rm\tilde v}_*}= [\vec{\rm V}(R)- \vec{V}(R_\odot)]+ [\vec{\rm v}_*-\vec{\rm v}_\odot] .
\end{equation}

We refer the reader to Mignard (2000) for the use of this equation in connection to the Galactic rotation model obtaining the star's peculiar velocity. For a given homogeneous group of stars, the Oort constants were obtained by this author, and a careful analysis led to determining the local peculiar velocity field. In the present work we do not intend to repeat this analysis, so we have opted to use his determination to recover the peculiar velocity field. Therefore by using the parallaxes and proper motions of the sampled stars, we could obtain the motion of the Sun and the peculiar velocities necessary to the  kinematical  analysis of the nearby disk stars. The components of the solar motion used here were extracted from Table 2 in Mignard (2000) for each interval of spectral type evaluated from the B-V color, as well as the Oort constants.

The peculiar velocity vector of a star in the Galaxy is determined by the components  $\vec{\rm u}$, $\vec{\rm v}$, and $\vec{\rm w}$, respectively directed toward the Galactic center ($\hat{\vec{\rm u}}$), the direction of positive increase of Galactic longitudes ($\hat{\vec{\rm v}}$), and the direction of the North Galactic Pole ($\hat{\vec{\rm w}}$). For the purposes of our present analysis, instead of using this general dynamic system, it is rather more convenient to adopt a reference frame projecting these components along the line of sight direction, $\vec{\rm v_r}$, and the two axes of Galactic coordinates, $\vec{\rm v_l}$ and $\vec{\rm v_b}$. Since we are dealing with nearby objects, there is no loss of information, and the conversion between these two right-hand coordinate systems is easily made by adopting the transformations

\begin{eqnarray}
{\rm u}=&{\rm v_r}\cos l \cos b &-{\rm v_b}\cos l\sin b -{\rm v_l}\sin l \nonumber\\
{\rm v}=&{\rm v_r} \sin l\cos b &-{\rm v_b} \sin l\sin b +{\rm v_l}\cos l\label{eq01}\\
{\rm w}=&{\rm v_r} \sin b &+{\rm v_b} \cos b . \nonumber 
\end{eqnarray}

\noindent These equations enable us to express the nearby Galactic velocity field in terms of more directly observable velocity components that could be inferred from the parallax and proper motion data. 

In general terms, the statistical distribution describing the Galactic structure of a group of stars in a specified unit volume is defined in such a way that the number of objects in the velocity space $d{\rm u}d{\rm v}d{\rm w}$ is 

\begin{equation}
\delta N = n f({\rm u,v,w}) d{\rm u} d{\rm v} d{\rm w}= n f'({\rm v}_r,{\rm v}_b,{\rm v}_l) d{\rm v}_rd{\rm v}_b d{\rm v}_l ,\label{eq02}
\end{equation}

\noindent where the symbol $n$ represents the local spatial density of stars, and the second identity is a direct consequence of the Jacobian of the above velocity transformations being equal to unity.  In the present analysis, we need to deal with another limitation: the radial velocities are not known, at least for the vast majority of stars in our sample. Therefore we are restricted to using the velocity field as deduced from the parallax and proper motion data, projected into the celestial sphere. For that reason it will be more convenient to work with the projected velocity distribution defined by

\begin{equation}
\delta N = n g({\rm v}_b,{\rm v}_l) d{\rm v}_bd{\rm v}_l ,\label{eq03}
\end{equation}

\noindent where the function

\begin{equation}
g({\rm v}_b,{\rm v}_l)=\int_{-\infty}^{+\infty} f'({\rm v}_r,{\rm v}_b,{\rm v}_l) d{\rm v}_r\label{eq04}
\end{equation}

\noindent represents the velocity distribution integrated along the line of sight of the observer. In writing this equation we implicitly assume that our sample faithfully represents the radial velocity distribution. In that context we can define our statistical problem as the determination of an expression for $g({\rm v}_b,{\rm v}_l)$ that could be  applied to describing the Hipparcos parallax and proper motion database.

As usual we adopt the view that the observed ellipsoid of velocities distribution of disk stars is aligned with the directions $\hat{\vec{ \rm u}}, \hat{\vec{\rm v}}, \hat{\vec{\rm w}}$ and is approximately described by the so-called one-component Scwarzschild distribution

\begin{equation}
f({\rm u,v,w}) = \frac{1}{(8\pi^3)^{1/2}\sigma_{\rm u}\sigma_{\rm v}\sigma_{\rm w}}\exp(Q)\label{eq05}
\end{equation}

\noindent where

\begin{equation}
Q=-\frac{1}{2}(\frac{{\rm u}^2}{\sigma_{\rm u}^2} +\frac{{\rm v}^2}{\sigma_{\rm v}^2} +\frac{{\rm w}^2}{\sigma_{\rm w}^2}) .\label{eq06}
\end{equation}

To ease the notation of the expressions below, we define the anisotropy parameters $\Gamma_{\rm v}=\sigma_{\rm u}/\sigma_{\rm v}$ and $\Gamma_{\rm w}=\sigma_{\rm u}/\sigma_{\rm w}$, describing the dispersions of the velocity ellipsoid in terms of the velocity dispersion towards the Galactic center. Moreover, we write the distribution function expressing the velocities in terms of the radial velocity dispersion ($ {\rm v}_r\equiv {\rm v}_r/\sigma_{\rm u},{\rm v}_l\equiv {\rm v}_l/\sigma_{\rm u}, {\rm v}_b\equiv {\rm v}_b/\sigma_{\rm u} $). By adopting these definitions, we conclude that the transformation of the Swarzschild ellipsoid to the projected velocity distribution system is simply given by the expression 

\begin{equation}
Q=-\frac{1}{2}c_{rr}{\rm v}_r^2 -\frac{1}{2}c_{ll}{\rm v}_l^2 -\frac{1}{2}c_{bb}{\rm v}_b^2
+c_{rl}{\rm v}_r{\rm v}_l +c_{rb}{\rm v}_r{\rm v}_b +c_{lb}{\rm v}_l{\rm v}_b , \label{eq07a}
\end{equation}

\noindent where the numerical coefficients are easily determined by a direct comparison of this equation with the result of adopting the transformations of Eq. \ref{eq01} into Eq.\ref{eq06}:

\begin{eqnarray}
c_{rr}=&\cos^2l\cos^2b +\Gamma_{\rm v}^2\sin^2l\cos^2b +\Gamma_{\rm w}^2\sin^2b\nonumber\\
c_{ll}=&\sin^2l +\Gamma_{\rm v}^2\cos^2l \nonumber\\
c_{bb}=&\cos^2l\sin^2b +\Gamma_{\rm v}^2\sin^2l\sin^2b +\Gamma_{\rm w}^2\cos^2b\label{eq07}\\
c_{rl}=&(1 -\Gamma_{\rm v}^2)\sin l\cos l \cos b \nonumber\\
c_{rb}=&(\cos^2l +\Gamma_{\rm v}^2\sin^2l  -\Gamma_{\rm w}^2)\sin b\cos b \nonumber\\
c_{lb}=&-(1 -\Gamma_{\rm v}^2)\sin l \cos l \sin b .\nonumber
\end{eqnarray}

Using these coefficients we can obtain the corresponding projected distribution of velocities by an integration along the radial direction,

\begin{eqnarray}
g({\rm v}_l,{\rm v}_b)=\frac{\Gamma_{\rm v}\Gamma_{\rm w}}{(8\pi^3)^{1/2}}   
\exp(-\frac{1}{2}c_{ll}{\rm v}_l^2 -\frac{1}{2}c_{bb}{\rm v}_b^2 +c_{lb}{\rm v}_l{\rm v}_b)\nonumber\\ 
\int_{-\infty}^{+\infty} \exp[-\frac{1}{2}c_{rr}{\rm v}_r^2 +(c_{rl}{\rm v}_l+c_{rb}{\rm v}_b){\rm v}_r] d{\rm v}_r . \label{eq08}
\end{eqnarray}

\noindent We can simplify this relation by using the the tabulated integral

\begin{displaymath}
\int_{-\infty}^{+\infty} \exp(-p^2 x^2 +qx) dx = \exp(\frac{q^2}{4p^2}) \frac{\sqrt{\pi}
}{p}
\end{displaymath}

\noindent to obtain the expression

\begin{eqnarray}
g(v_l,v_b)=\frac{\Gamma_v\Gamma_w}{(4\pi^2 c_{rr})^{1/2}}\nonumber \;\;\;\;\;\;\;\;\;\;\;\;\;\;\;\;\;\;\;\;\;\;\;\;\;\;\;\;\;\;\;\; \;\;\;\;\;\;\;\;\;\;\;\;\;\;\;\;\;\;\;\;\;\; \hbox{} \\
\exp[-\frac{1}{2}c_{ll}v_l^2 -\frac{1}{2}c_{bb}v_b^2 +c_{lb}v_lv_b 
+\frac{(c_{rl}v_l+c_{rb}v_b)^2}{2c_{rr}}]\nonumber
\end{eqnarray}

\noindent and, collecting the coeficients corresponding to the same power of $v_l$ and $v_b$, we obtain the equation

\begin{equation}
g({\rm v}_l,{\rm v}_b)=\frac{\Gamma_{\rm v}\Gamma_{\rm w}}{(4\pi^2 c_{rr})^{1/2}}   
\exp[-\frac{1}{2}q_{ll} {\rm v}_l^2 -\frac{1}{2}q_{bb} {\rm v}_b^2 +q_{lb}{\rm v}_l{\rm v}_b] \label{eq09} 
\end{equation}

\noindent where the numerical coefficients are given by

\begin{eqnarray}
q_{ll}=\frac{\Gamma_v^2\cos^2b +\Gamma_w^2\sin^2l\sin^2b +\Gamma_{\rm v}^2\Gamma_{\rm w}^2\cos^2l\sin^2b}
{\cos^2l\cos^2b+\Gamma_{\rm v}^2\sin^2l\cos^2b +\Gamma_{\rm w}^2\sin^2b} \nonumber\\
q_{bb}=\frac{\Gamma_{\rm w}^2(\cos^2l +\Gamma_{\rm v}^2\sin^2l)}
{\cos^2l\cos^2b+\Gamma_{\rm v}^2\sin^2l\cos^2b +\Gamma_{\rm w}^2\sin^2b}\label{eq10}\\
q_{lb}=\frac{(\Gamma_{\rm v}^2-1)\Gamma_{\rm w}^2\sin l\cos l \sin b}
{\cos^2l\cos^2b+\Gamma_{\rm v}^2\sin^2l\cos^2b +\Gamma_{\rm w}^2\sin^2b} . \nonumber 
\end{eqnarray}

Depending on the choice of the observing direction, we obtain some particularly interesting cases for this projected velocity distribution. For example, in the direction $b=0$, we are observing the Galactic plane and therefore ${\rm v}_b={\rm w}$. In that case the parameters describing the velocity distribution are $q_{ll}=\Gamma_{\rm v}^2/(\cos^2l +\Gamma_{\rm v}^2\sin^2l)$, $q_{bb}=\Gamma_{\rm w}^2$  and $q_{lb}=0$. Therefore the component of the velocity distribution along the latitude direction ${\rm v}_b$ is identical to the distribution of the velocity ellipsoid in the direction orthogonal to the Galactic plane (${\rm w}$). Standing in the Galactic plane but looking now towards $l=0$, the longitude velocity component (${\rm v}_l$) is equal to the velocity component in the rotation direction ${\rm v}$, while we have ${\rm v}_l={-\rm u}$ by looking at $l=\pi/2$.

Another case of interest occurs by looking towards $b=\pi/2$ at the North Galactic Pole. In that region $q_{ll}=\sin^2l +\Gamma_{\rm v}^2\cos^2l$, $q_{bb}=\cos^2l+ \Gamma_{\rm v}^2\sin^2l$, and $q_{lb}=(\Gamma_{\rm v}^2-1)\sin l\cos l$. If we approximate the pole by keeping  $l=0$, we have $\rm{v}_b=-{\rm u}$, ${\rm v}_l={\rm v}$. And if $l=\pi/2$ we have ${\rm v}_b=-{\rm v}$, ${\rm v}_l=-{\rm u}$ and $q_{ll}=1$, $q_{bb}=\Gamma_{\rm v}^2$. 

In principle, the bivariate distribution function $g({\rm v}_l,{\rm v}_b)$ obtained can be directly compared with the observations at each direction on the celestial sphere. Since in each region the contribution of the fundamental vectors ${ \vec{\rm u}, \vec{\rm v}, \vec{\rm w}}$ changes with the Galactic coordinates, we should be able to deduce their influence in the distribution function by comparing the prediction of this distribution with a large database. But unfortunately the number of stars in our sample is not that great for justifying the use of a fine grid, and we instead prefer to work with the marginal distributions $s({\rm v}_l)=\int g({\rm v}_l,{\rm v}_b) d{\rm v}_b$ and $t({\rm v}_b)=\int g({\rm v}_l,{\rm v}_b) d{\rm v}_l$ and compare these distributions in some selected sectors of the sky. Using Eq. \ref{eq09} for the projected distribution, we can easily obtain

\begin{equation}
s({\rm v}_l)=\frac{1}{\sqrt{2\pi\sigma_l}} \exp(-\frac{{\rm v}_l^2}{2\sigma_l^2}),\label{eq11}
\end{equation}

\noindent showing that the marginal distribution in Galactic longitude is Gaussian with a dispersion

\begin{equation}
\sigma_l^2=(\frac{\cos^2l}{\Gamma_{\rm v}^2}+ sin^2l)\sigma_{\rm u}^2 \label{eq12}
\end{equation}

\noindent that continously changes over the celestial sphere. Analogously, the marginal velocity distribution along the Galactic latitude is

\begin{equation}
t({\rm v}_b)=\frac{1}{\sqrt{2\pi\sigma_b}} \exp(-\frac{{\rm v}_b^2}{2\sigma_b^2})\label{eq13}
\end{equation}

\noindent where

\begin{equation}
\sigma_b^2=(\cos^2l\sin^2b+ \frac{\sin^2l\sin^2b}{\Gamma_v^2}+ \frac{\cos^2b}{\Gamma_w^2}) \sigma_u^2 .\label{eq14}
\end{equation}

Therefore the two marginal velocity distributions are also Gaussian, but their dispersions vary with the direction of observation. The extension of these expressions to a two-component fluid system is straightforward, resulting in

\begin{equation}
\delta N = n [\alpha g_1({\rm v}_b,{\rm v}_l) + (1-\alpha)g_2({\rm v}_b,{\rm v}_l)] d{\rm v}_bd{\rm v}_l ,\label{eq15}
\end{equation}

\noindent and the corresponding marginal distributions are 

\begin{eqnarray}
s({\rm v}_l)=\alpha s_1({\rm v}_l) +(1-\alpha)s_2({\rm v}_l) \nonumber\\
t({\rm v}_b)=\alpha t_1({\rm v}_b) +(1-\alpha)t_2({\rm v}_b) \label{eq16}  
\end{eqnarray}

\noindent where the dispersions of each component are given by Eqs. \ref{eq12} and \ref{eq14}, and the parameter $\alpha$ is a measure of the relative importance of the contribution of each component. 

\section{Analysis}

Since the dispersions involved in the marginal distributions depend upon combinations of square sine and cosine of $l$ and $b$, we chose to compare these distributions with the observational data by dividing the celestial sphere in 72 equal-area regions reduced to nine regions in the first octant of the sphere ($0\leq l\leq 90, 0\leq b\leq 90$). For practical purposes, we adopted the three longitude segments located at 0-30$^o$, 30$^o$-60$^o$, and 60$^o$-90$^o$. The corresponding latitude segments were defined at $b=0$, $b=\arcsin(1/3)\simeq  19^o.5$,  and $b=\arcsin(2/3)\simeq 41^o.8$. By adopting that coarse division, each spherical sector has the same area, keeping a reasonable number of objects of 200-300 in each region. Inside each region we binned the ${\rm v}_l, {\rm v}_b$ data in 50 velocity intervals having $\Delta {\rm v} =4 {\rm km.s}^{-1}$, therefore spanning the range of $-100, +100 {\rm km.s}^{-1}$, which is adequate for studying the kinematics of the disk stars.  

%                                             Simple A&A Table
%_____________________________________________________________
%
\begin{table*}
\caption{Kinematical parameters}             % title of Table
\label{tab1}      % is used to refer this table in the text
\centering                          % used for centering table
\begin{tabular}{c c c c c c c c c}        % centered columns (4 columns)
\hline\hline                 % inserts double horizontal lines
Sample& N& $<\delta {\rm N}^2>^{1/2}$ &$\chi^2/df$& $\alpha$&$\sigma_u$&$\Gamma_v$&$\Gamma_w$\\
\hline                        % inserts single horizontal line
A0A5&4202&1.22$\pm$0.04& 0.74&0.78$\pm$0.06 &19.54$\pm$1.60& 1.63$\pm$0.33& 2.74$\pm$0.30\\
    &    &             &     &             & 5.88$\pm$1.26& 0.69$\pm$0.29& 2.45$\pm$1.18 \\
    &    &             &     &             &             &         &          \\
    &    &1.36$\pm$0.05& 0.80&---      & 15.25$\pm$1.07& 1.35$\pm$0.21& 2.65$\pm$0.30 \\    
\hline                                   %inserts single line
A5F0&3185&0.95$\pm$0.03&0.57&0.53$\pm$0.10& 24.85$\pm$2.88& 1.64$\pm$0.46& 2.58$\pm$0.59\\
    &    &             &    &             & 15.18$\pm$1.60& 1.37$\pm$0.31& 3.48$\pm$0.86 \\
    &    &             &    &             &             &          &          \\
    &    &0.99$\pm$0.03&0.57&---       & 19.45$\pm$1.22& 1.49$\pm$0.20& 2.99$\pm$0.38 \\    
\hline      
F0F5&2837&0.94$\pm$0.03&0.56&0.77$\pm$0.20&24.50$\pm$2.25&1.59$\pm$0.28&2.19$\pm$0.43\\
    &    &             &    &             & 19.37$\pm$4.57& 1.72$\pm$0.84& 7.47$\pm$2.20 \\
    &    &             &    &             &              &         &          \\
    &    &0.96$\pm$0.03&0.55&---          & 23.07$\pm$1.62& 1.59$\pm$0.22& 2.66$\pm$0.44 \\    
\hline          
K0K5&6533&1.01$\pm$0.03&1.18&0.57$\pm$0.07&40.66$\pm$4.48&1.49$\pm$0.37&2.07$\pm$0.51\\
    &    &             &    &             & 23.11$\pm$2.31& 1.70$\pm$0.38& 4.00$\pm$1.12 \\
    &    &             &    &             &          &             &          \\
    &    &1.19$\pm$0.04&1.49&---          & 31.37$\pm$1.60& 1.60$\pm$0.22& 2.84$\pm$0.47 \\    
\hline              
K5M0&3350&0.99$\pm$0.03&0.99&0.65$\pm$0.09&41.25$\pm$5.19&1.54$\pm$0.47&2.07$\pm$0.59\\
    &    &             &    &             & 20.02$\pm$3.33& 1.23$\pm$0.41& 3.80$\pm$1.97 \\
    &    &             &    &             &          &             &          \\
    &    &1.09$\pm$0.04&1.12&---          & 31.57$\pm$2.84& 1.42$\pm$0.26& 2.56$\pm$0.54 \\    
\hline                  
M0M5&2285&0.94$\pm$0.03&0.87&0.74$\pm$0.08&41.28$\pm$5.20&1.32$\pm$0.36&2.37$\pm$0.70\\
    &    &             &    &             & 18.54$\pm$4.34& 2.04$\pm$1.42& 3.67$\pm$2.28 \\
    &    &             &    &             &          &             &          \\
    &    &1.06$\pm$0.04&1.04&---          & 33.10$\pm$3.50& 1.43$\pm$0.33& 2.70$\pm$0.65 \\    
\hline     
\end{tabular}
\end{table*}

A minimization code was projected to evaluate the sum of the square deviation of the marginal distributions $s(\sigma_l,\sigma_{\rm u},\Gamma_{\rm v},\Gamma_{\rm w};l,b)$, $t(\sigma_b,\sigma_u,\Gamma_v,\Gamma_w;l,b)$. In each sector  we consider that the dispersions $\sigma_l=\sigma_l(\bar{l},\bar{b})$ and $\sigma_b=\sigma_b(\bar{l},\bar{b})$ are constant, where $\bar{l},\bar{b}$ were evaluated at the centers of the corresponding spherical sectors. Then, for each spectral type, the routine minimizes the combined mean square deviation of all the ${\rm v}_l$ and ${\rm v}_b$ histograms in the nine pre-defined celestial regions. Therefore the total number of bins is $n_{bin}=2x50x9=900$, and this database was used to estimate the mean square deviation.

In Table \ref{tab1} we present a summary of the determination of all the relevant kinematical parameters by the two ways: single and double Gaussian velocity distributions. For each spectral class, we have three rows of parameters. The first two rows indicate the minimization result using the two-component population model. The third row contains a determination of the parameters based on a single kinematical population. In column $N$, we indicate the total number of stars in each sample of spectral types, while $<\delta N^2>^{1/2}$ indicates the corresponding mean square deviation and its respective error estimated at a two-sigma level. In the next column we present the final chi-square per degree of freedom ($\chi^2/df$). To estimate the chi-square, we evaluated the error at each velocity bin containing $n$ objects as its Poisson expectation $\sqrt{n}$. In the other columns we have the proportion $\alpha$ in each case, the velocity dispersion in the $\hat{\vec{\rm u}}$ direction and the anisotropy parameters. For each parameter we also give the error bar evaluated at a two-sigma level.

From the inspection of this table we can observe that, for early type stars (A0-F5), there is no significant net gain in describing the kinematical data with the two-population model. For these objects the variation in the mean square deviation between the two models is close to the expected intrinsic statistical fluctuation of their mean deviation. The same conclusion is reached when we compare the $\chi^2/df$ of the one- and two-component models. This one-component disk population can be directly compared with the results of Bienaym\'e (\cite{Bienayme99}) who obtained an average value of $\sigma_u \simeq  21.3 {\rm km.s}^{-1}$ for the early stars group, $\sigma_v \simeq  11.3 {\rm km.s}^{-1}$ ($\Gamma_v=1.9$),$\sigma_w \simeq  9.0 {\rm km.s}^{-1}$ ($\Gamma_w=2.4$). Similar values were also obtained by Mignard (\cite{Mignard}) who have determined $\sigma_u = 19.8 {\rm km.s}^{-1}$, $\sigma_v =13.3 {\rm km.s}^{-1} (\Gamma_v =1.5)$, $\sigma_w =8.4 {\rm km.s}^{-1} (\Gamma_w =2.4)$. 

\begin{figure*}
 \centering
 \includegraphics[width=\textwidth]{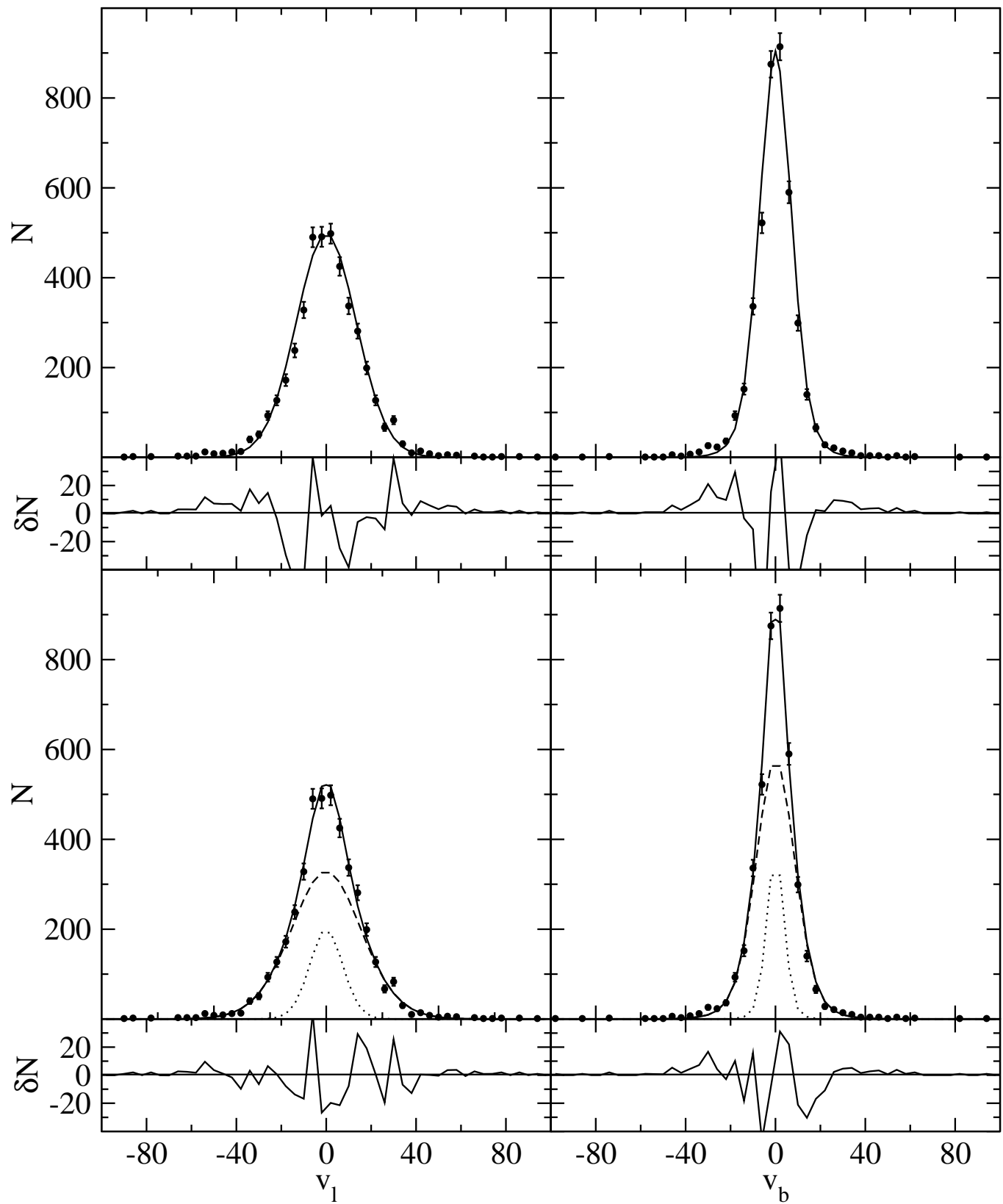}
 \caption{Distribution of the longitude, left panels, and latitude, right panels, components of velocity. The two upper panels show the result of fitting a single population of the Scwarzschild velocity distribution. In the lower panels we present the result of fitting a two-fluid population. Below each panel, we show the residual distribution of the fit. }\label{ModA0A5}
\end{figure*}

\begin{figure*}
 \centering
 \includegraphics[width=\textwidth]{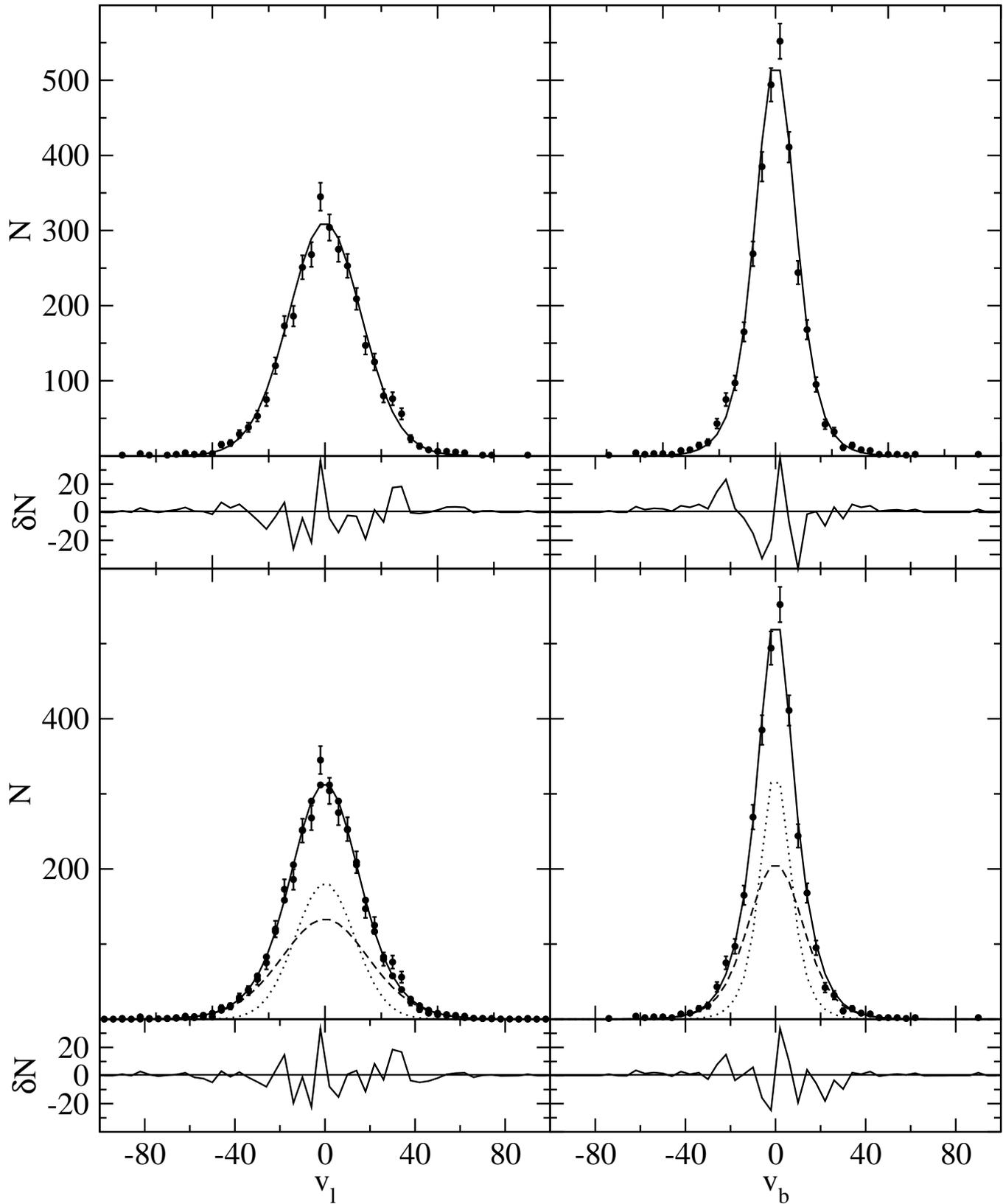}
 \caption{Same as Fig. \ref{ModA0A5} for the A5F0 population.}\label{ModA5F0}
\end{figure*}

\begin{figure*}
 \centering
 \includegraphics[width=\textwidth]{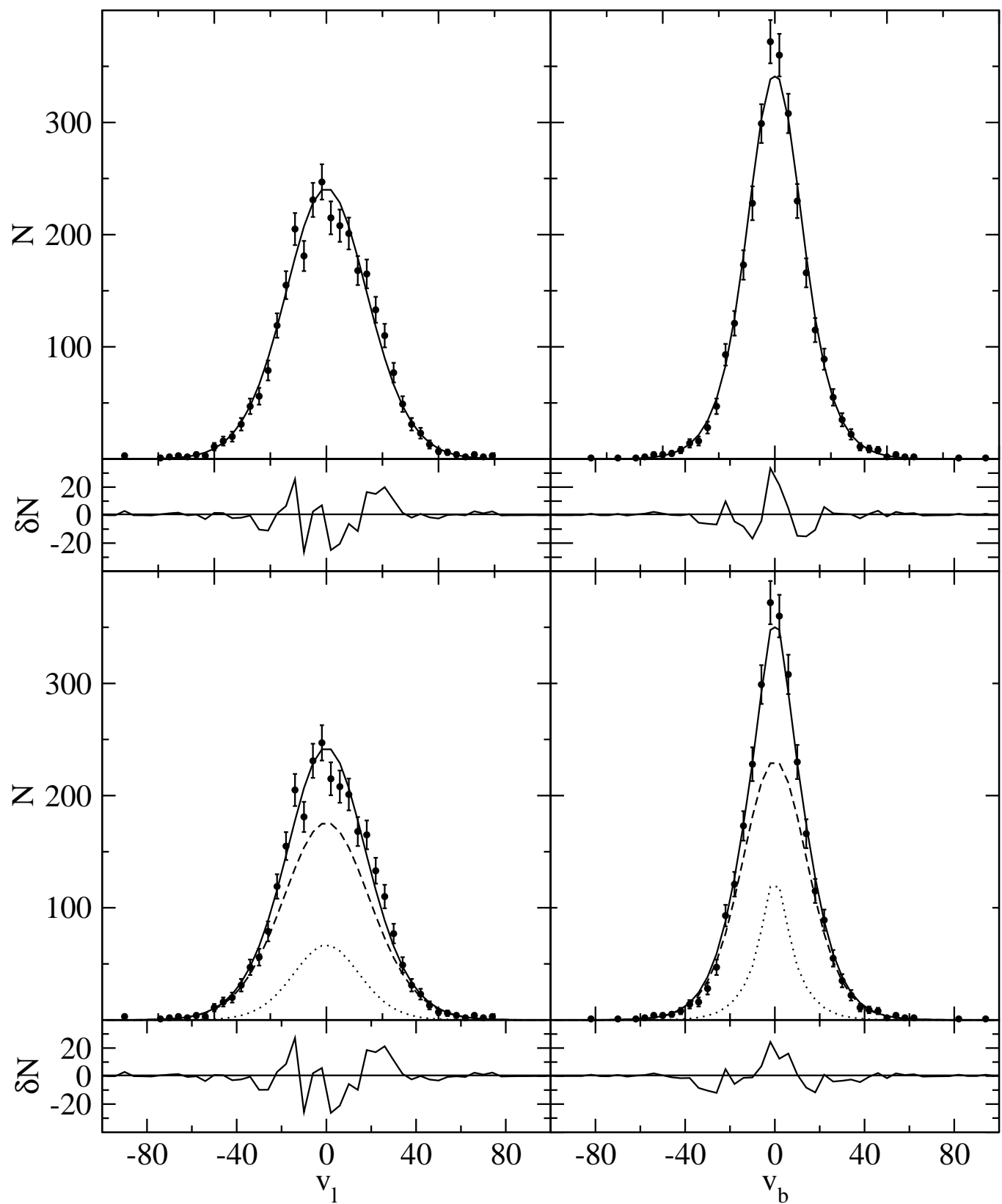}
 \caption{Same as Fig. \ref{ModA0A5} for the F0F5 population}\label{ModF0F5}
\end{figure*}

In Figs. \ref{ModA0A5}, \ref{ModA5F0}, and \ref{ModF0F5} we present the distribution of peculiar velocities for this group of early type stars in graphical terms.   The top diagrams in these plots represent the results of the one-population model, while the lower panels shows the corresponding two-population model. The left panel represents the longitude velocity distribution, while the right panel represents the latitude velocity distribution. In both sets of panels, there is a small plot of the residual best model fit. In order to simplify the visual inspection we present these plots corresponding to the result of adding all the distributions for each spherical sector in a single one. But we point out that the minimization code takes all the individually selected sectors into account since, as we saw above, the velocity dispersion varies with their position in the celestial sphere. In all these three groups corresponding to the early type stars, we can observe that the gain from considering a two-fluid population is barely marginal. In particular the residual distribution has basically the same structure in all these cases, implying that there is no gain in considering a two-population model description. A more quantitative estimate was done using the Wald-Wolfowitz test to verify if the number of sign changes in the residue distribution is compatible with the null hypothesis of them being normally distributed. For the early type stars, there is no case where we could reject simultaneously both the $v_l$ and $v_b$ distributions to the 95\% confidence level.  

The more uncertain case among early type stars occurs in the A0-A5 class where the improvement in the fit ($\delta(<\delta {\rm N}^2>^{1/2}) \simeq 0.14$) is barely larger than twice the error bar of the single-component solution ($\varepsilon=0.05$).Using the Wald-Wolfowitz test, we could reject the hypothesis that the residual $v_l$ distribution is randomly distributed at a 95\% confidence level but not $v_b$. Moreover the oscillatory behavior seen in Fig. \ref{ModA0A5} is removed when a  low velocity dispersion population is added, giving some support to its reality. The effect is probable real, but we need a larger sample to be more positive about its detection. 

For the late type stars, our single population model basically, as expected, recovers the same distribution of late type stars as found by Bienaym\'e (\cite{Bienayme99}): $\sigma_u \simeq  34.4 {\rm km.s}^{-1}$, $\sigma_v \simeq  23.8 {\rm km.s}^{-1}$ ($\Gamma_v\simeq 1.4$), $\sigma_w \simeq  17.7 {\rm km.s}^{-1}$ ($\Gamma_w\simeq 1.9$). And similar values were also found by Mignard (\cite{Mignard}), who estimated $\sigma_u = 31.2 {\rm km.s}^{-1}$, $\sigma_v=22.0 {\rm km.s}^{-1} (\Gamma_v=1.4)$, $\sigma_w=17.2 {\rm km.s}^{-1} (\Gamma_w=1.8)$. But contrary to what we found in early type stars, in the present case of the late type giant stars, there is an obvious improvement in the mean square deviation ($<\delta {\rm N}^2>^{1/2}$) by considering the two-fluid model description. For each spectral class, the improvement is greater than three times the mean expected error fluctuation, as we can verify in Table \ref{tab1}. In the one-population model we use the Wald-Wolfowitz test to reject the null hypothesis that the residual velocities are randomly distributed at the 95\% confidence level in both $v_l$ and $v_b$ distributions in all the three spectral classes. The same type of improvement can be seen by the $\chi^2/df$, which is systematic better in the two-fluid model. Therefore we feel that there is a real gain in describing this population by a two-component model. In gross terms the kinematics of late type giants can be described by a superposition of two well defined populations. The low velocity dispersion component has an average dispersion ellipsoid ($\sigma_u, \Gamma_v,\Gamma_w= 21.1\pm 3.3, 1.7\pm.7, 3.8\pm 1.2$) that is similar to the average single-population solution found in early type stars ($\sigma_u, \Gamma_v,\Gamma_w=19.3\pm 1.3, 1.5\pm 0.2, 2.8\pm 0.4$). Therefore it is tempting to interpret that these low velocity dispersion populations as having the same origin as the thin disk population of early type stars. 

\begin{figure*}
 \centering
 \includegraphics[width=\textwidth]{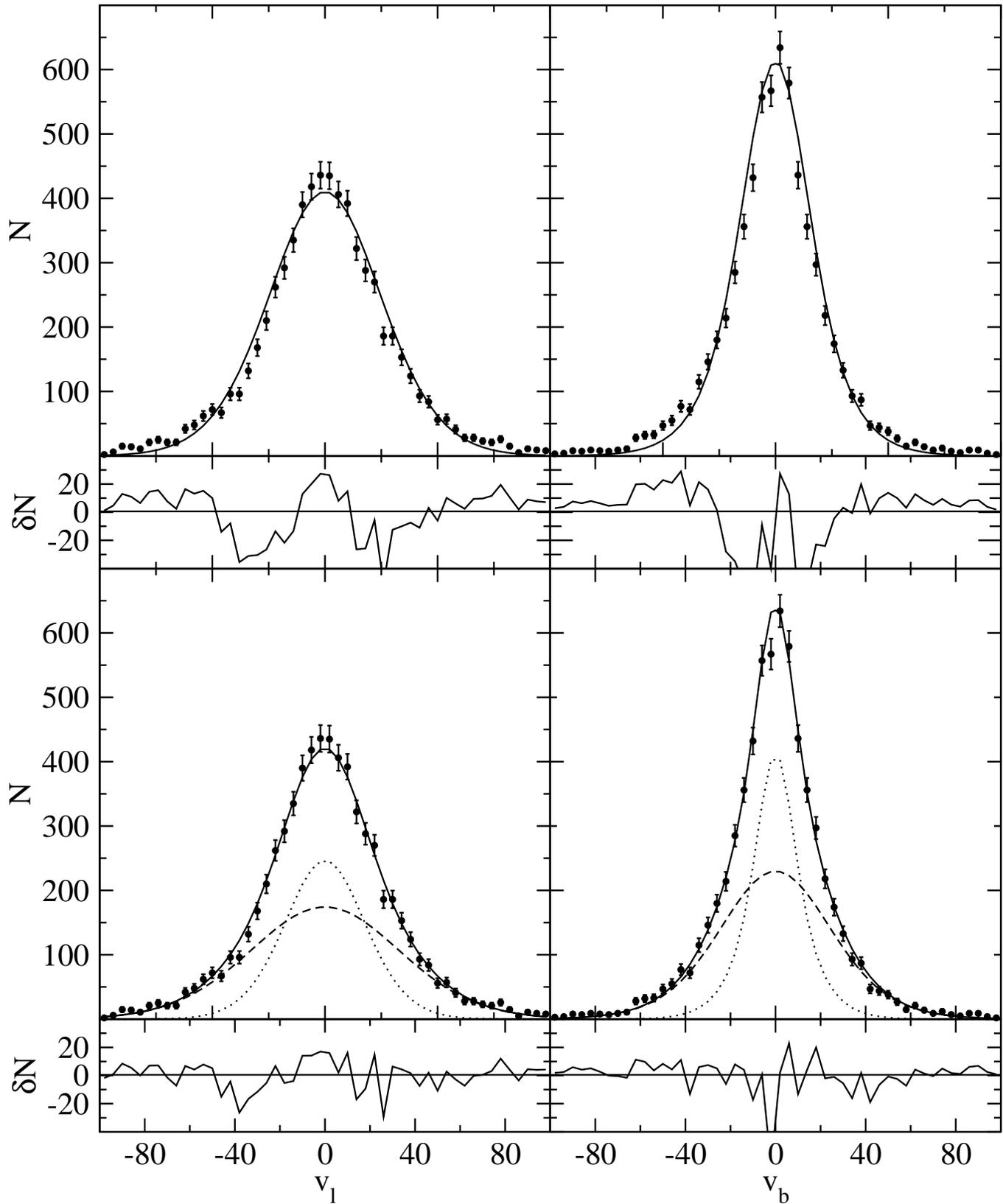}
 \caption{Same of Fig. \ref{ModA0A5} for the K0K5 population.}\label{ModK0K5}
\end{figure*}

\begin{figure*}
 \centering
 \includegraphics[width=\textwidth]{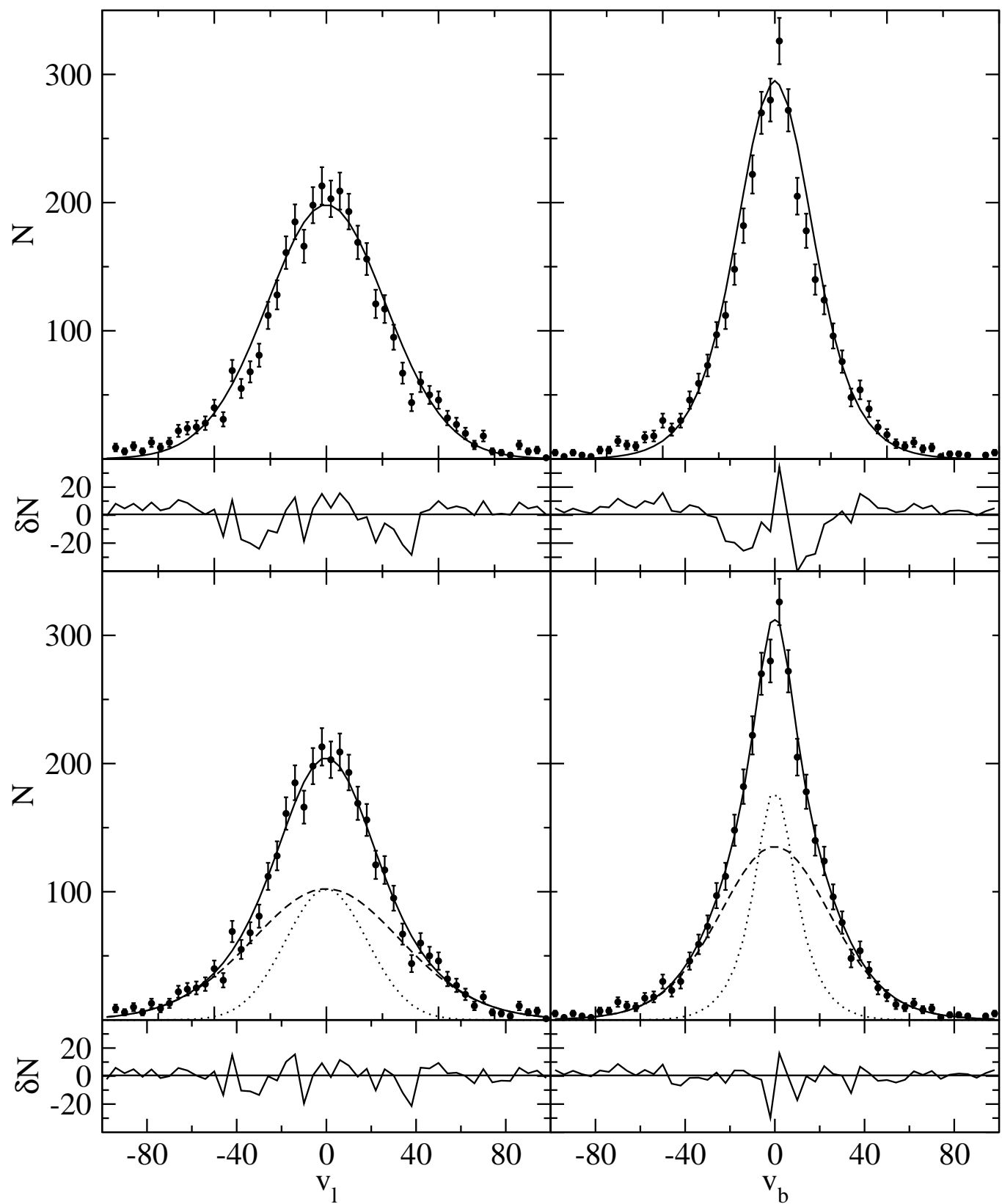}
 \caption{Same as Fig. \ref{ModA0A5} for the K5M0 population.}\label{ModK5M0}
\end{figure*}

\begin{figure*}
 \centering
 \includegraphics[width=\textwidth]{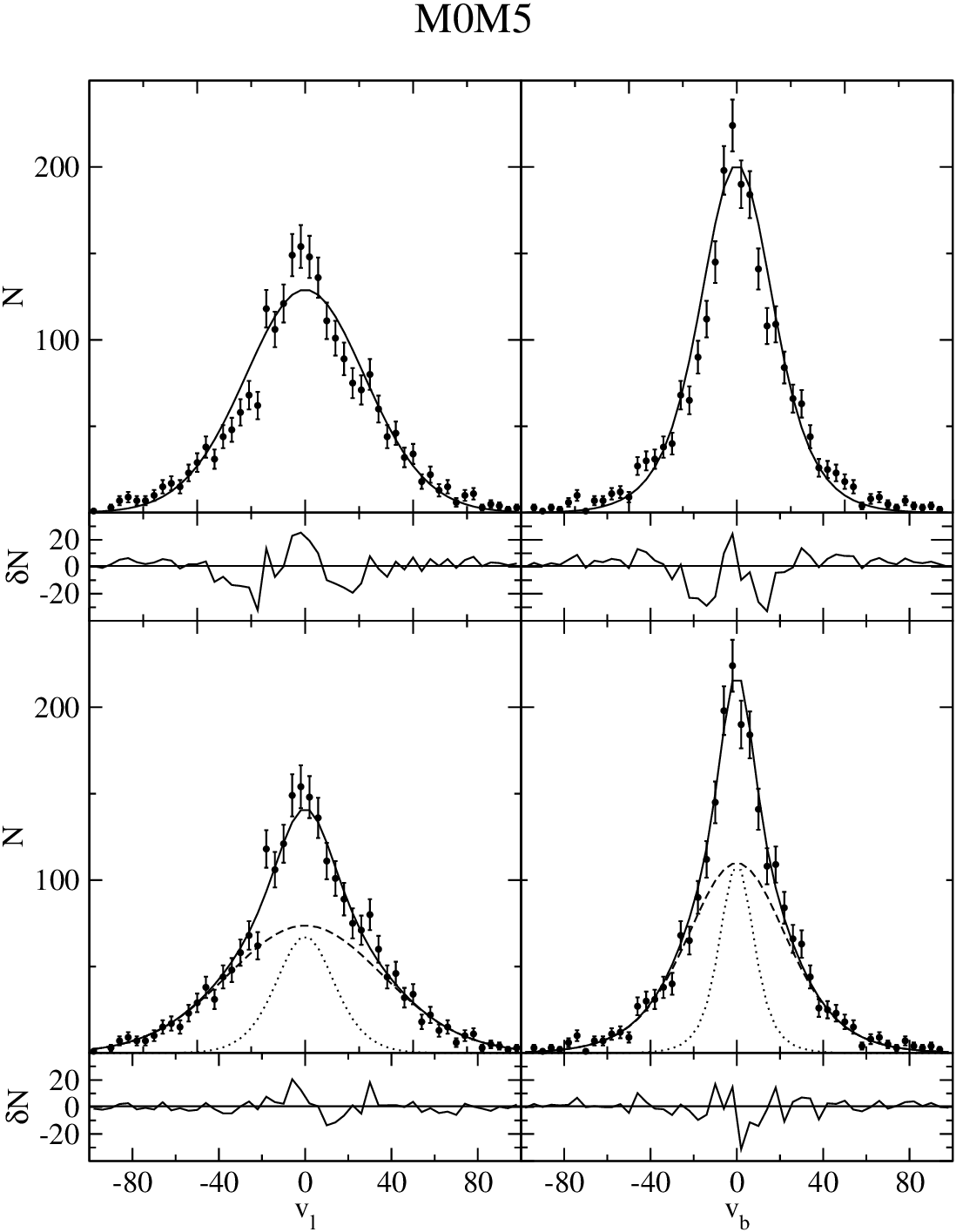}
 \caption{Same as Fig. \ref{ModA0A5} for the M0M5 population.}\label{ModM0M5}
\end{figure*}

The high-velocity component of late type stars has a distinct average ellipsoid:( $\sigma_u, \Gamma_v,\Gamma_w=41.0\pm 4.9, 1.5\pm.0.4, 2.2\pm 0.6$) when compared to the early type population. The gain in the statistical description can also be clearly appreciated in Figs. \ref{ModK0K5}, \ref{ModK5M0}, and \ref{ModM0M5}. We observe from these plots that the single model population has the same distinct oscillatory residual structure corresponding to a depression in the high-velocity wings of the distribution and an excess in the low-velocity center. This occurs because the use of one single population is clearly not able to simultaneously reproduce the wings as well as the central low-velocity regime of the distribution. In the high-velocity wings, this one population model slightly underestimates the observed data points, while the contrary occurs in the central region. 

The final result is that the minimum residual solution in this model results from a compromise trying to describe the low-velocity and high-velocity regions at the same time. This result is the oscillatory effect seen in the residual plots of the upper panels in these diagrams. The poorest results are obtained in bins where the residual reaches values in excess of $20$ objects.  Clearly the algorithm tries to find a compromise but is unable to deliver a randomly distributed error structure. In contrast, the consideration of a two-fluid population seen in the lower panels removes this oscillatory behavior, since the code uses one population to describe the high-velocity wings and another one to describe the central low-velocity core. That is therefore why the final adjustment seen in Table \ref{tab1} gives a much more satisfactory fit in the wings and also in the central region of low-velocity stars. The mean proportion of high-velocity dispersion stars, measured by the $\alpha$ parameter, corresponds to approximately 65\% of the objects, while the remaining 35\% is composed of low dispersion stars. It is quite interesting that this rough proportion is maintained in all three spectral types with no clear trend toward systematic variation. 

In crude terms these two kinematical population found in the late type stars should have distinct scale heights above the Galactic plane in order to sustain their hydrostatic equilibrium in the gravitational field. In a multicomponent disk, the scale length of a given component is approximately given by $z_0=\sigma^2_W/\pi G\Sigma_{od}$, where $\Sigma_{od}$ represents the total projected mass density on the disk (Bahcall \& Casertano, \cite{Bahcall84}).  We note that the total mass density in the Galactic disk includes not only the stelar component but also the unseen contribution of the dark matter halo and other baryonic components such as gas and dust. 

The exact share of these components in the whole mixture is a matter of debate in the literature. According to Cr\'ez\'e et al (\cite{Creze98}), the local dynamical mass deduced from Hipparcos data is $\rho_0=0.076\pm0.015 {\rm M_\odot/pc^3}$, and the total projected mass density should be close to $\Sigma_{0d}=33 {M_\odot/pc^2}$, a lower value than previous estimates. Therefore, adopting this estimation, the two kinematical population found above should have a logarithmic scale length of $z_0= 54$ pc and $z_0= 403$ pc. The exact figure of these estimations is naturally subject to confirmation of the local disk density of matter. In Fig. \ref{DensZType} we have indicated the prediction corresponding to this estimative by the two continuous lines. In the case of early type stars, the estimation based on this model fits the observed density distribution observed from our sample quite nicely. For the late type giants, the observed distribution deviates from this prediction probably because these stars are feeling the more extended disk potential that cannot be adequately be described by the thin disk approximation. But in both cases, it is clear that the two kinematical populations of late type stars can be considered as members of the Galactic thin disk.
 
When we consider the whole range of spectral types, another interesting point is that the mean velocity dispersion, as deduced from the single population model, steadily increases by almost a factor of two when we consider the progression from the early type to the late type stars. As can be seen from Table \ref{tab1}, this progression of the velocity dispersion is quite smooth as we go from A0-A5 towards the M0-M5 class. But we can also see from our two-fluid model that the internal velocity dispersions of each late type population remains almost constant. Therefore the reason for the continous increase in the velocity dispersion is that the contribution of these two populations changes as we go from the K0-K5 to the M0-M5 populations. In the former case 57\% of the stars belong to the high-velocity population, while almost 74\% are high-velocity stars in the later case. Thus, even if the velocity dispersion of the  populations of low-velocity and high-velocity stars remains approximately constant, their combined dispersions increase continously depending on their proportion.

\begin{figure*}
 \centering
 \includegraphics[width=\textwidth]{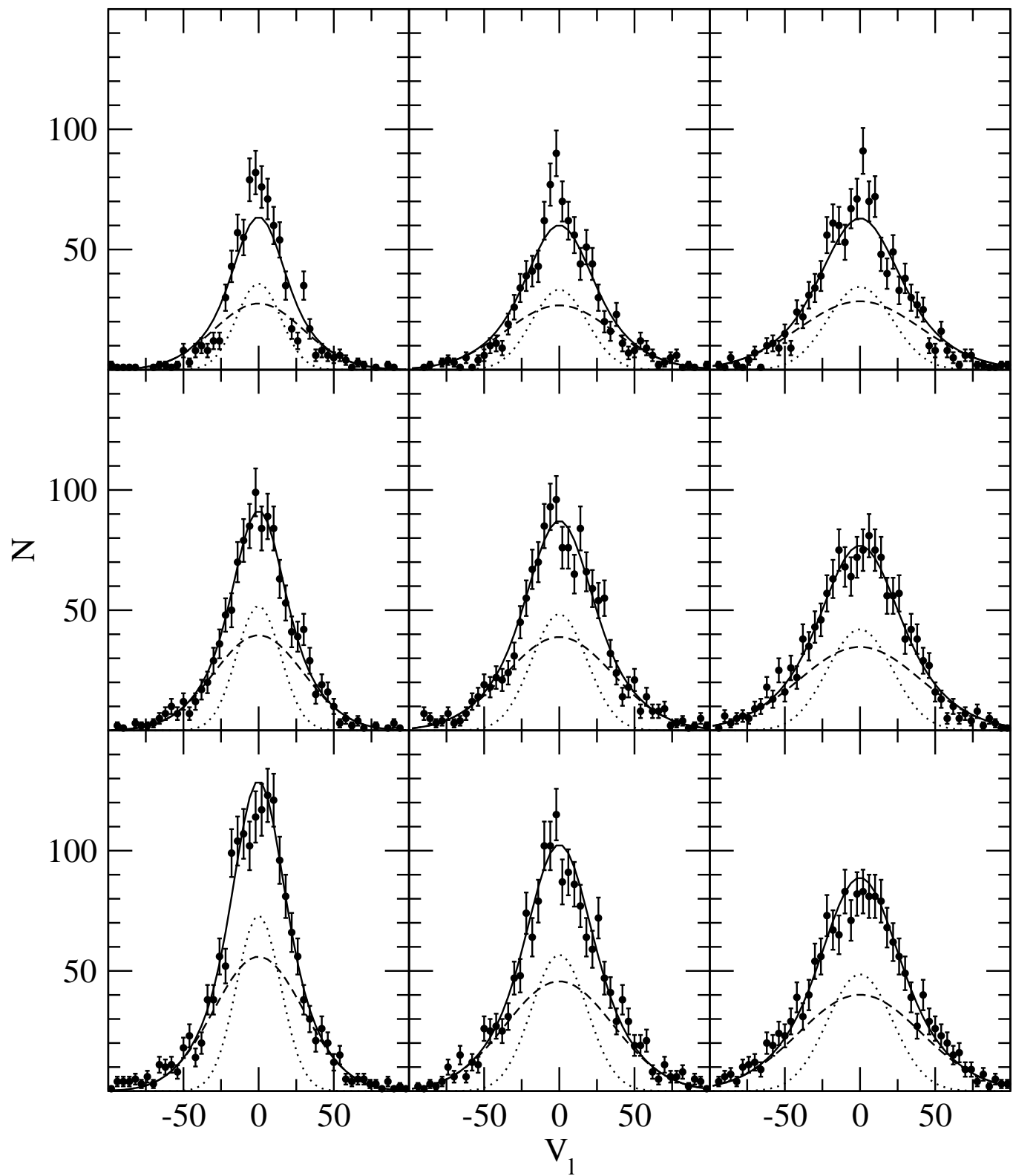}
 \caption{Comparison of the longitude velocity distribution of the two models with the combined data of K0K5+K5M0+M0M5 late type stars. The panels correspond to the nine sectors in the celestial sphere seen in Galactic coordinates. The three upper panels correspond to the the Galactic plane ($\bar{b}\simeq 9^o.8$), the second row are the intermediate zone ($\bar{b}\simeq 30^o.7$),  and the lower row corresponds to the galactic pole regions ($\bar{b}\simeq65^o.9$). The three columns correspond to the longitude zones centered on  $\bar{l}=15, 45, 75$.  }\label{VlSector}
\end{figure*}

In fact, the two populations described above have quite similar properties to the dispersions of low and high metalicity stars studied by Str\"omgren (\cite{Stromgren87}). From this work the author concludes that the population of high-metallicity stars $0<[Fe/H]<-0.4$ has a mean dispersion $\sigma_u \simeq 20 {\rm km.s}^{-1}$, while the dispersion for low-metallicity $-0.4<[Fe/H]<-0.8$  steadily increases reaching  $\sigma_u \simeq 40 {\rm km.s}^{-1}$ at the low-metallicity extreme. 

\begin{figure*}
 \centering
 \includegraphics[width=\textwidth]{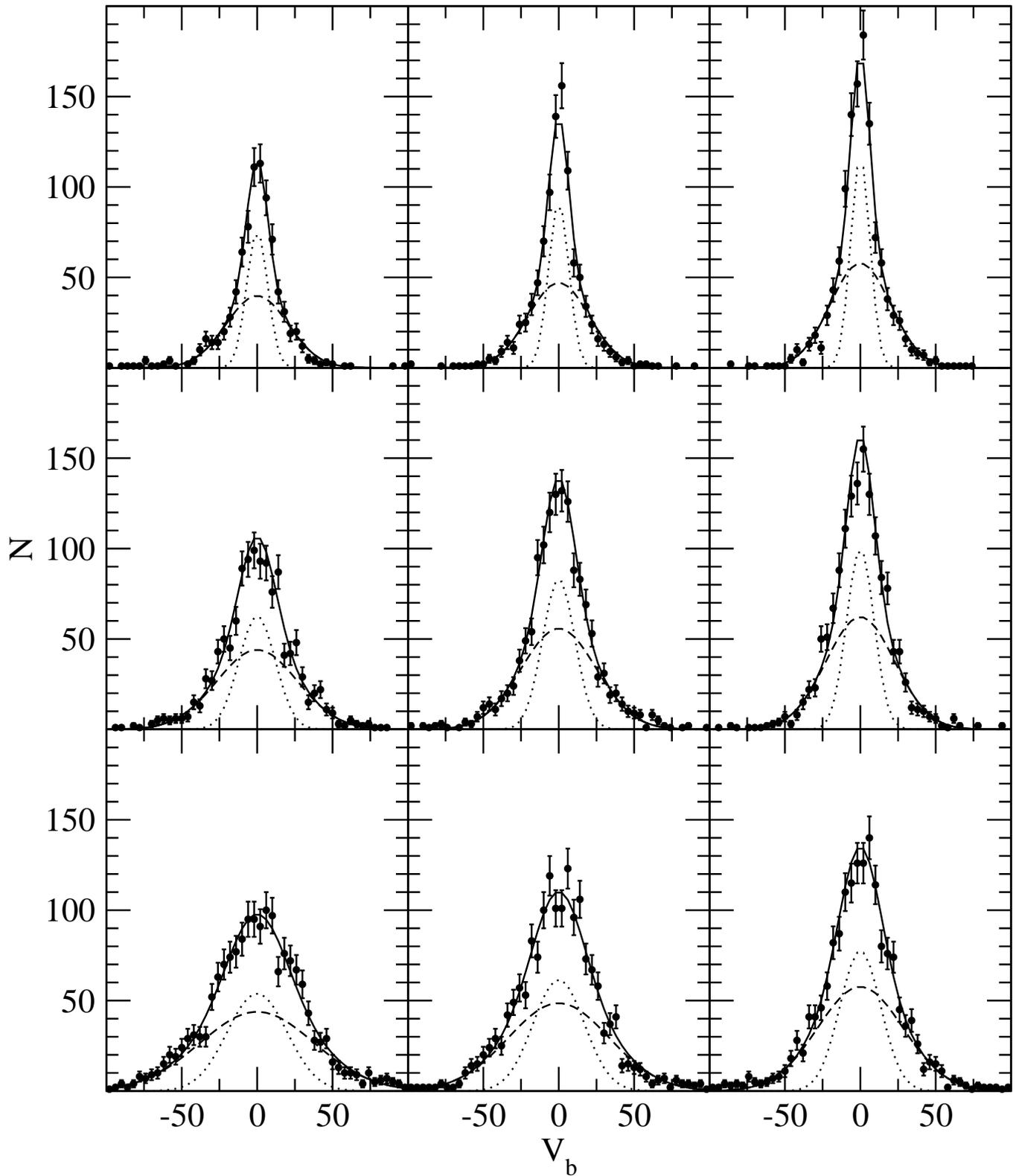}
 \caption{Same as Fig. \ref{VlSector} for the latitude velocity component.}\label{VbSector}
\end{figure*}

In Figs. \ref{VlSector} and \ref{VbSector}, we show the results for the two-population model applied to the combined class of the K0K5+K5M0+M0M5 sample. The ${\rm v}_b$ distribution in Fig. \ref{VbSector} shows a very good agreement of the model for all nine sectors. No systematic effect is present and the model represent all the sectors equally well. The ${\rm v}_l$ distribution shown in Fig. \ref{VlSector} has a small systematic effect in the first three upper panels corresponding  to the directions close to the galactic plane ($\bar{b}\simeq 9.^o8$). The effect implies an excess of low-velocity stars in the database  as compared with the prediction. In the wings of these three panels, we can also see a slightly oscillatory perturbation when the data points are compared to the model. Probably this effect is caused by a residual due to the subtraction of the galactic rotational velocity that might be different for the two kinematical populations. The first upper panel of this figure should have a dispersion close to $\sigma_{\rm v}$, so at first glance, it would imply an excess of low-velocity stars in the direction of the Galactic rotation. But if that were the case, we should expect the same effect to be present in the right lower panel of Fig. \ref{VbSector} and that is not the case. Therefore the possible origin of this excess is not clear to us.
    
\section{Conclusions}

Our analysis has shown that the kinematical properties of the population of early type stars (A0-F5) is described well by a single Swarzschild velocity distribution. No net gain is really detected when we use a two-population model even considering the large number of objects in the present sample. Probably the detection of an additional minor population, if it exists, would require a much larger sample and a greater accuracy than the present one in the Hipparcos database, as expected by the Gaia experiment. There might be a hint of the presence of another small population since the residual plot on Figs. \ref{VlSector} and \ref{VbSector} presents some structure in the low-velocity stars. But the size of the actual population was not large enough to permit a clear detection for our analysis. 

In the case of late type stars, the situation is clearly defined and points to the presence of two distinct kinematical populations. One population has high velocity dispersion stars with $\sigma_{\rm u}\simeq 40 {\rm km.s}^{-1}$, while the low velocity dispersion stars have $\sigma_{\rm u}\simeq 20 {\rm km.s}^{-1}$. The low-velocity component in the giant branch, corresponding to approximately 35\% of this population, is typical of the thin disk kinematics. According to Caloi et al. (\cite{Caloi99} see their Fig. 2), the youngest stars presently reaching the giant branch have ages in the range $5 {\rm x} 10^8 - 1 {\rm x} 10^9{\rm yr}$ ,corresponding to objects with ages comparable to a few Galactic rotational periods. Therefore is tempting to identify these low velocity dispersion giants with thin disk stars that are now evolving towards the giant branch. In that case the progression from the early type main sequence stars to these thin disk giants shows practically no variation in their velocity dispersion. In fact our result is consistent with a single population of  thin disk objects having $\sigma_u, \Gamma_v, \Gamma_w=20.2\pm 2.3, 1.6\pm.5$, and $3.3\pm 1.1$.

The high-velocity late type giant population is typical of a more extended disk  as discussed in several other analyses. It is quite improbable that our high-velocity sample could represent a transition to the thick disk. Gould et al. (\cite{Gould96}) found that the solar neighborhood contains 80\% of stars belonging to a disk having a scale length of 326 pc, while  the other 20\% of the population belongs to an extended disk with a scale length of 656 pc. Since we are sampling a scale length well inside this thin disk, the degree of contamination by the more extended thick disk should be even lower than this figure. According to Soubiran (\cite{Soubiran93}), this thin, old disk population is characterized by $\sigma_U=44\pm 6 {\rm km.s}^{-1}$ and  $\sigma_V=24\pm 4 {\rm km.s}^{-1}$, close to the solution we found for our higher velocity giants. On the contrary, the thick disk would require a higher velocity dispersion of $\sigma_U=56\pm 11 {\rm km.s}^{-1}$ and  $\sigma_V=43\pm 6 {\rm km.s}^{-1}$. Although we cannot discard the presence of some degree of contamination, our feeling is thus that our high velocity dispersion late type population belongs to the thin disk. 

Another piece of information comes from the work of Caloi et al. (\cite{Caloi99}), where they study a large sample, including both giants and main sequence stars, using Hipparcos parallaxes and radial velocities extracted from the literature. An interesting point raised by their discussion is that stars having low space velocity relative to the LSR ($-40 {\rm km.s}^{-1}< V<10 {\rm km.s}^{-1} $) tend to present a higher proportion of young objects (age$< 10^7 {yr}$). On the other hand, the bulk of objects in their sample having velocities comparable to the high-velocity tails of our sample ($-60 {\rm km.s}^{-1}< V<-40 {\rm km.s}^{-1} $) appear older than $ 10^9 {yr}$. By increasing the velocity range, they begin to sample the transition zone to the thick disk and the inner halo objects. Therefore our present result supports the idea that our high-velocity population is due to local disk stars having ages comparable with a few rotation periods of the Galaxy. In that case the dynamical differentiation detected in this work could probably reflect the diffusion of stellar orbits due to the dynamical relaxation due to the irregular  time-varying gravitational field in the Galactic disk (Wielen \cite{Wielen77}). In particular, Griv et al. (\cite{Griv01}) estimate that scattering by  potential irregularities in the galactic disk could cause a velocity variation in $10^9 {\rm yr}$ comparable to the difference we have detected here. Therefore, in this scenario, the older objects would correspond to the population of the high-velocity tail of our late giant distribution.

There is however one point that might challenge this interpretation, since our high-velocity population present no sign of variation among the spectral classes K0-K5, K5-M0, and M0-M5. In all of these three samples the results are remarkably consistent with a single kinematical population that is quite distinct from the one found in the low-velocity thin disk. There is no sign of a smooth transition between the two regimes, which would be expected in the presence of a gradual migration phenomenon such as the scattering by irregularities. In that case one could naively expect that the kinematical properties would smoothly change as we sample objects with different classes and probably different ages and chemical composition. That is not what we see since the kinematical properties of our sample of late type giants are quite consistent with a single kinematical population. In that context it is interesting to note the work of Norris ( \cite{Norris87}), according to which the prevalence of the scattering mechanism implies a smooth transition regime, whereas the presence of two overlapping distributions would favor a discrete-origin model. In fact, Carney et al. \cite{Carney89} use this argument in favor of the existence of a discrete population in the disk with [m/H] $\simeq -0.5$ and a velocity dispersion perpendicular to the disk of $40 {\rm km}·{\rm s}^{-1}$. It is suggested that this population could be the signature of an early merger event occurring shortly after the disk had formed.

\begin{acknowledgements}
Part of this work was supported by FAPESP project number 00/06695-0.
\end{acknowledgements}

\end{document}